\documentclass[%
,onecolumn%
,secnumarabic%
,amssymb, amsmath,nobibnotes, aps, prb]{revtex4}

\usepackage{bm}%
\usepackage[dvips]{graphicx, color}

\usepackage{comment}
 
\expandafter\ifx\csname package@font\endcsname\relax\else
 \expandafter\expandafter
 \expandafter\usepackage
 \expandafter\expandafter
 \expandafter{\csname package@font\endcsname}%
\fi

\newcommand{\Tii}{Ti$_{1-x}$Co$_x$O$_{2}$}
\begin{document}
\title{Bulk and Surface Magnetization of Co atoms in Rutile Ti$_{1-x}$Co$_x$O$_{2-\delta}$ Thin Films Revealed by X-Ray Magnetic Circular Dichroism}%

\author{V.~R.~Singh}
\email{vijayraj@wyvern.phys.s.u-tokyo.ac.jp}
\affiliation{Department of Physics, University of Tokyo, 
Bunkyo-ku, Tokyo 113-0033, Japan}

\author{Y.~Sakamoto}
\affiliation{Department of Physics, University of Tokyo, 
Bunkyo-ku, Tokyo 113-0033, Japan}

\author{T.~Kataoka}
\affiliation{Department of Physics, University of Tokyo, 
Bunkyo-ku, Tokyo 113-0033, Japan}

\author{M.~Kobayashi}
\affiliation{Department of Physics, University of Tokyo, 
Bunkyo-ku, Tokyo 113-0033, Japan}

\author{Y.~Yamazaki}
\affiliation{Department of Physics, University of Tokyo, 
Bunkyo-ku, Tokyo 113-0033, Japan}

\author{A.~Fujimori}
\affiliation{Department of Physics, University of Tokyo, 
Bunkyo-ku, Tokyo 113-0033, Japan}

\author{F.-H.~Chang}
\affiliation{National Synchrotron Radiation Research Center (NSRRC), Hsinchu 30076, Taiwan, Republic of China}

\author{D.-J.~Huang}
\affiliation{National Synchrotron Radiation Research Center (NSRRC), Hsinchu 30076, Taiwan, Republic of China}

\author{H.-J.~Lin}
\affiliation{National Synchrotron Radiation Research Center (NSRRC), Hsinchu 30076, Taiwan, Republic of China}

\author{C.~T.~Chen}
\affiliation{National Synchrotron Radiation Research Center (NSRRC), Hsinchu 30076, Taiwan, Republic of China}

\author{H.~Toyosaki}
\affiliation{Institute for Materials Research, Tohoku University, 
Sendai 980-8577, Japan}

\author{T.~Fukumura}
\affiliation{Institute for Materials Research, Tohoku University, 
Sendai 980-8577, Japan}
\affiliation{PRESTO, Japan Science and Technology Agency, Kawaguchi 332-0012, Japan}

\author{M.~Kawasaki}
\affiliation{Institute for Materials Research, Tohoku University, 
Sendai 980-8577, Japan}
\affiliation{WPI Advanced Institute for Materials Research, Tohoku University, 
Sendai 980-8577, Japan}
\affiliation{CREST, Japan Science and Technology Agency, Kawaguchi 332-0012, Japan}

\date{\today}

\begin{abstract}
We have studied magnetism in Ti$_{1-x}$Co$_x$O$_{2-\delta}$ thin films with various $x$ and $\delta$ by soft x-ray magnetic circular dichroism (XMCD) measurements at the Co $L_{2,3}$ absorption edges. The estimated ferromagnetic moment by XMCD was 0.15-0.24 $\mu_B$/Co in the surface, while in the bulk it was 0.82-2.25 $\mu_B$/Co, which is in the same range as the saturation magnetization of 1.0-1.5 $\mu_B$/Co. These results suggest that the intrinsic origin of the ferromagnetism. The smaller moment of Co atom at surface is an indication of a magnetically dead layer of a few nm thick at the surface of the thin films.
\end{abstract}


\maketitle

In the rapidly growing research field known as ``spintronics'', there is nowadays considerable interest in the search for dilute magnetic semiconductor (DMS) with high Curie temperatures ($T_C$), with  particular focus on the exploration of different semiconductor hosts. Among them, Ti$_{1-x}$Co$_x$O$_{2-\delta}$ has attracted much  attention \cite{YMR,YM,HT,MS,WT,JY,SR,SRS,RRA}. Co-doped anatase TiO$_2$ thin films were first reported to be ferromagnetic even above room temperature ($\sim$ 400K) \cite{YMR} with a magnetic moment of 0.32 $\mu_B$/Co, and subsequenct works \cite{YM} reproduced the  room temperature ferromagnetism also in rutile Ti$_{1-x}$Co$_x$O$_{2-\delta}$, and the magnetic ordering was attributed to carrier-induced ferromagnetism \cite{HT} as in the III-V based DMS \cite{JK}. In these works, it was  concluded that  the Co ions substitute for the Ti cations and that Ti$_{1-x}$Co$_x$O$_{2-\delta}$ can be viewed as DMS \cite{KME}. Some experimental studies have reported that $T_C$ is up to $\sim$ 650 K \cite{RIKH}. However, a large spread of $T_C$ and magnetization has been kept to be reported, depending on the preparation conditions, the resultant oxygen content, and the distribution and concentration of Co atoms (for review see Ref. \cite{RJ}). Anomalous Hall effect (AHE) measurements on rutile  Ti$_{1-x}$Co$_x$O$_{2-\delta}$  films with systematically different Co contents and carrier doping levels \cite{HT} have confirmed the room temperature ferromagnetism in this material and strongly suggest that the ferromagnetism is intrinsic and not due to Co clustering. At the same time as the AHE measurements were reported, a magnetization study of highly reduced rutile thin film sample revealed the co-occurrence of superparamagnetism and AHE, casting a doubt on the reliability of  AHE for determining whether the ferromagnetism in Ti$_{1-x}$Co$_x$O$_{2-\delta}$   is intrinsic or extrinsic \cite{SRS}, and the origin of ferromagnetism still remain unclear.

In this Letter, we report measurements of x-ray absorption spectroscopy (XAS) and x-ray magnetic circular dicroism (XMCD) at the Co $L_{2,3}$ edges of rutile Ti$_{1-x}$Co$_x$O$_{2-\delta}$ thin films with various Co concentration $x$ and oxygen vacancy $\delta$  to clarify the origin of the ferromagnetism. XAS is well known as a very useful method to determine the electronic structure of transition-metal atoms and its XMCD signal reflects the magnetic state of the same atoms. Mamiya {\it et al.'}s XMCD spectra of Ti$_{1-x}$Co$_x$O$_{2-\delta}$ \cite{KME} are distinctly different from the previous results by Kim {\it et al.} \cite{JY}, where they studied Ti$_{1-x}$Co$_x$O$_{2}$  samples heat-treated prior to the measurements. They observed no multiplet features at the Co  $L_{2,3}$ edges in their XMCD spectra, attributing  the XMCD signals to those of segregated metallic Co clusters. The segregated Co clusters in their samples obviously arose from high annealing temperature (673K) in a vacuum, as indicated by the systematic increase of the XAS and XMCD signals of metallic Co \cite{JY}, and may not be due to the intrinsic properties of \Tii. While Mamiya {\it et al.'}s Co $L_{2,3}$ result \cite{KME} clearly indicates that the Co$^{2+}$ ions in rutile Ti$_{1-x}$Co$_x$O$_{2-\delta}$ are the divalent high-spin ionic state, Co$^{2+}$, indicating that the doped Co  atoms  substituted for the Ti sites and are responsible for the ferromagnetism \cite{KME}. However the XMCD signal corresponds to only $\sim$ 0.1 $\mu_B$/Co, much smaller than the saturation magnetization of $\sim$ 1.0 $\mu_B$/Co deduced from SQUID measurements for bulk \cite{TFEA} . In this study, we compare XMCD results taken using the surface sensitive in agreement with Mamiya {\it et al.} \cite{KME} total electron yield (TEY) and those taken using the bulk-sensitive total fluorescence yield (TFY) mode, together with SQUID measurements. In the TEY mode, we observed a small XMCD signal at the Co $L_{2,3}$ edges, which corresponds to $\sim$ 0.15-0.24 $\mu_B$/Co  while in the TFY mode it was 0.82-2.25 $\mu_B$/Co which is much larger than Mamiya {\it et al.}'s results \cite{KME} and consistent with SQUID measurements. We have thus found that the Co ion indeed has a large moment corresponding to the bulk magnetization. The spectral line shape can be accounted for by that of  Co$^{2+}$  atom, representing that the origin of magnetization should be Co atoms in the Ti$_{1-x}$Co$_x$O$_{2-\delta}$ oxide matrix.

Rutile Ti$_{1-x}$Co$_x$O$_{2-\delta}$ (101) epitaxial thin films with $x$ = 0.03, 0.05 and 0.10 were synthesized by the pulsed laser deposition method on r-sapphire substrates at 673K at different oxygen pressures,  $P_{{\rm O}_2}$= $10^{-6}$ or $10^{-7}$ Torr \cite{HT}. The samples fabricated in oxygen pressure $P_{{\rm O}_2}$=$10^{-6}$ and  $10^{-7}$ Torr are named as low-$\delta$ and high-$\delta$, respectively, since the number of oxygen vacancies increases with decreasing oxygen pressure.  Segregation of secondary phases were not observed under careful inspection by x-ray diffraction, AFM, scanning electron microscopy (SEM), and transmission electron microscopy (TEM). Its ferromagnetism at room temperature was confirmed by Hall-effect measurements, magnetization measurements, and MCD measurements in the visible region \cite{HT,TFEA,HTTF}. The XAS and XMCD measurements were performed at BL-11A of National Synchrotron Radiation Research Center, Taiwan. In XMCD measurements, magnetic fields was applied to the sample along out-of-plane. XAS and XMCD spectra were obtained in the TEY  and TFY modes without surface preparation in order to avoid  possible destruction of the sample surfaces. The probing  depth of the TEY mode and TFY mode was $\sim$ 5 and 100 nm, respectively. 


\begin{figure}[htbp]
\begin{center}
\includegraphics[width=08cm]{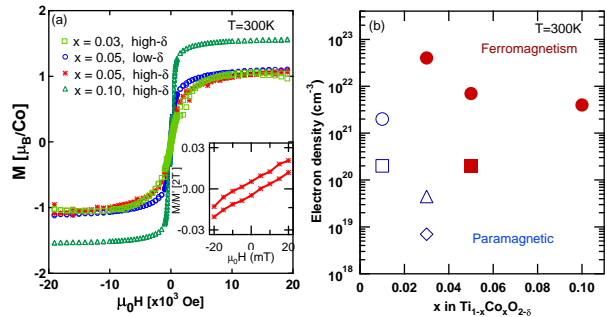}
\caption{(Color online)(a) Magnetization {\it vs.} magnetic field curves of rutile-type Ti$_{1-x}$Co$_x$O$_{2-\delta}$.Inset shows M (H) curve around zero magnetic field at 300 K. (b) Magnetic phase diagram as a function of electron carrier density $n_e$ and Co concentration deduced from ordinary Hall effect at 300 K for Ti$_{1-x}$Co$_x$O$_{2-\delta}$. Solid and open symbols denote ferromagnetic and paramagnetic samples, respectively. Circle, square, triangle and diamond symbols correspond to $P_{{\rm O}_2}$=$10^{-7}$, $10^{-6}$, $10^{-5}$ and $10^{-4}$ Torr, respectively, during synthesis.}
\end{center}
\end{figure}
Figure 1(a)-(b) shows magnetic properties  of Ti$_{1-x}$Co$_x$O$_{2-\delta}$ at 300K with different {\it x} and electron carrier density ($n_e$). It is clear from Fig.1 that the magnetization M(H) was in the range 1.0-1.5 $\mu_B$/Co with coercive force around several tens of Oersted, and increases with $\delta$ or $n_e$. In M(H) measurements, magnetic field was applied to the sample along out-of-plane i.e. direction of rutile (101). AHE measurements  for Ti$_{1-x}$Co$_x$O$_{2-\delta}$  with different $n_e$ and {\it x} also showed the same magnetic field dependences. The resultant magnetic ``phase diagram'' shows that higher  $n_e$ and {\it x} induce the ferromagnetic phase as shown in Fig.1(b).
\begin{figure}[htbp]
\begin{center}
\includegraphics[width=08cm]{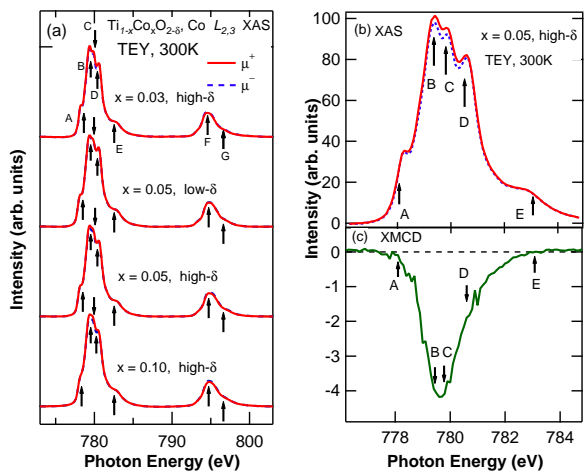}
\caption{(Color online) (a) Co $L_{2,3}$  XAS  and (b)-(c) XAS and XMCD  spectra of rutile-type Ti$_{1-x}$Co$_x$O$_{2-\delta}$ samples for \textsl{x}=0.05, high-$\delta$  taken in the TEY mode at \textsl {T} = 300 K and \textsl {H} = 1 T.}
\end{center}
\end{figure}

In Figure 2, we show the Co  $L_{2,3}$-edge XAS and XMCD spectra of Ti$_{1-x}$Co$_x$O$_{2-\delta}$  ({\it x}=0.03, 0.05 and 0.10 with low- and  high-$\delta$) thin films taken in the TEY mode. In the figure, $\mu_+$ and $\mu_-$ stand for the absorption coefficients for photon helicity parallel and antiparallel to the Co majority spin direction, respectively. The XMCD spectra ($\Delta\mu$= $\mu_+$ - $\mu_-$) have been corrected for the degree of circular polarization of the incident light. The XAS spectra of the rutile-type Ti$_{1-x}$Co$_x$O$_{2-\delta}$  thin films showed multiplet features. Here, we follow Mamiya {\it et al.} \cite{KME} and refer to each multiplet feature as A-G. The XMCD spectra show clear multiplet features that correspond almost one-to-one to those in the XAS spectra. The line shapes of the XAS and XMCD spectra are almost the same as Mamiya {\it et al.} \cite{KME}. The estimated magnetic moment is in the range of the 0.15-0.24 $\mu_B$/Co consistent with Mamiya {\it et al.} \cite{KME}, while the saturation moments deduced from the SQUID magnetization measurements is 1.0-1.2 $\mu_B$/Co. In contrast to Kim {\it et al.} \cite{JY}, the present experiment clearly revealed multiplet features in the XMCD spectra corresponding to those in  XAS without  annealing, consistent with the ferromagnetism arising from Co$^{2+}$ ions which are coordinated by O$^{2-}$ ions \cite{KME}. The experimental data, i.e., XAS and XMCD spectra, show qualitatively the good agreement with the calculated spectra for the Co$^{2+}$ high-spin configuration in the $D_{2h}$ crystal field \cite{KME}.

\begin{figure}[htbp]
\begin{center}
\includegraphics[width=08cm]{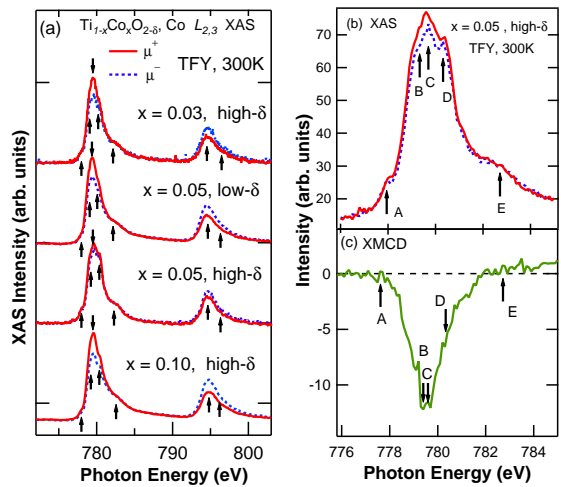}
\caption{(Color online) (a) Co $L_{2,3}$  XAS  and (b)-(c) XAS and XMCD  spectra of rutile-type Ti$_{1-x}$Co$_x$O$_{2-\delta}$ samples for \textsl{x}=0.05, high-$\delta$ taken in the the TFY mode at \textsl {T} = 300 K and \textsl {H} = 1 T.}
\end{center}
\end{figure}

Figure 3 show the results of the Co $L_{2,3}$ XAS and XMCD spectra of the same samples taken in the TFY mode. From the figure, it is clear that the XMCD intensities are much higher than those taken in the TEY mode. The large difference between the bulk-sensitive TFY and surface-sensitive TEY modes indicates that there is a magnetically dead layer of $\sim$ 5 nm, at the surfaces of the samples, which is consistent with recent measurements of the film thickness dependence of AHE \cite{MNAK}. The spectral line shapes of the XAS and XMCD spectra of the {\it x} = 0.05, high-$\delta$ sample shows a clear multiplet feature while the other sample shows relatively weak multiplet features. A possible origin of the weakness of the multiplet features might be the limited S/N ratio. The fine structure, which is indicative of Co$^{2+}$, is more pronounced in the TEY than in the TFY modes. This is because in this compound, there should be at least as many oxygen vacancies as Co and electrostatic interaction from the oxygen vacancies may affect Co position significantly \cite{TMEA} and hence we incorporated the random crystal field in our calculations. In the surface region, which is observed by the TEY mode, the position of Co atoms may be optimized in a similar way because of the oxidation and less structural constraint at the surface, resulting in relatively uniform crystal field. On the other hand, in the bulk, probed by the TFY mode, the position of Co atoms might be frozen in various local structures. The random crystal fields in the 3D crystal lattice make the TFY spectra broad and different from the TEY spectra. The XAS and XMCD spectra taken in the TFY mode also show good agreement with the calculated spectra for the Co$^{2+}$ in bulk Ti$_{1-x}$Co$_x$O$_2$ with random crystal fields\cite {CMC}.The parameters used in the calculations listed in Table I.
 \begin{figure}[htbp]
\begin{center}
\includegraphics[width=08cm]{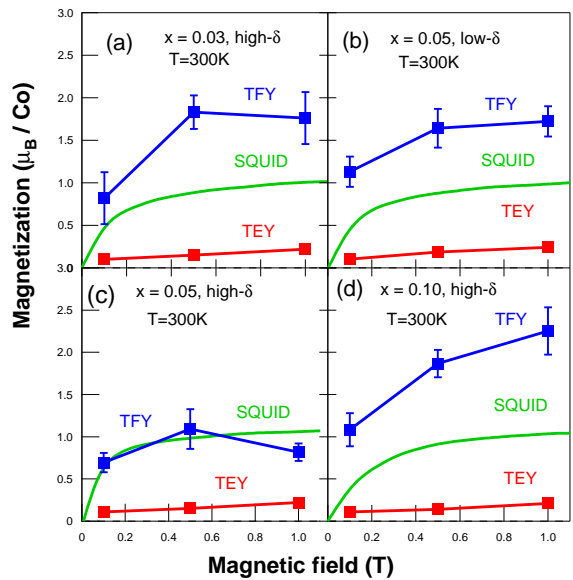}
\caption{(Color online) \textsl{M-H} relation of SQUID magnetization and magnetization estimated from the XMCD spectra Ti$_{1-x}$Co$_x$O$_{2-\delta}$.}
\end{center}
\end{figure}

\begin{table}[ht]
\caption{Electronic structure parameters for rutile Co-doped TiO$_2$ thin film used in the cluster-model calculations in units of eV : the charge-transfer energy $\Delta$, the on-site 3$d$-3$d$ Coulomb energy $U_{dd}$, and the 3$d$-2$p$ Coulomb energy $U_{dc}$ on the Co ion, the hopping integral between Co 3$d$ and O 2$p$ $V_{{\rm E}_g}$, and the crystal-field 10{\it Dq}.} 
\centering 
\begin{tabular}{c c c c c c c} 
\hline 
Crystal symmetry & $\Delta$ & $U_{dd}$ & $U_{dc}$ & $V_{{\rm E}_g}$  & 10{\it Dq} & weight(\%) \\ [0.5ex] 
\hline 
$D_{2h}$ low spin & 4 & 5 & 7 & 1.1 & 1.1$-$1.2 & 38\%\\ 
$O_h$ low spin & 3 & 6 & 7.5 & 1.1 & 1.1$-$1.2 & 38\%\\ [1ex] 
$O_h$ high spin & 2 & 5 & 7.5 & 1.1& 0.8$-$0.9  & 24\%\\ [1ex] 
\hline 
\end{tabular}
\label{table:nonlin} 
\end{table}

Figure 4  shows the magnetization estimated from the XMCD spectra taken in the TFY and TEY modes using optical sum rules \cite {CTCHEN} compared it with \textsl{M-H} curves from magnetization measurements as well as cluster model calculations. For the validity of sum rule in this case we divided the obtained spin-magnetic moment by a correction factor 0.92 \cite{YT}. The magnetic moment obtained from cluster model calculation is 1.48$\mu_B$/Co. Nevertheless, the Co magnetic moment is found to be obviously much larger in the bulk region than in the surface region. Since the TFY suffers from self-absorption so the magnetic moment obtained by optical sum-rule is not so accurate like TEY mode. The magnetization estimated from the XMCD spectra taken in the bulk-sensitive TFY mode are similar to those estimated from the SQUID measurements as well as cluster model calculations, which strongly suggest that the Co ions in the bulk region are responsible for the ferromagnetism while the surface layer of the film looks like magnetically dead layer, as confirmed by the XMCD taken in the surface-sensitive TEY mode. TiO$_2$ has an extraordinary chemical stability, hence we can rule out possible surface degradation as a cause of decrease in surface magnetization. From surface characterization techniques such as AFM and reflection high energy electron diffraction, we have not observed any change in the surface state. Also, from spectroscopic techniques, we have not observed a significant time dependence of XMCD and x-ray photoemission spectroscopy. Figure 1(a) shows magnetic hysteresis of Ti$_{1-x}$Co$_x$O$_{2-\delta}$ with different {\it x}. The coercive force is so small that the hysteresis is difficult to resolve by XMCD setup. The magnetic anisotropy with out-of-plane easy axis is not so strong in Ti$_{1-x}$Co$_x$O$_{2-\delta}$. As reported by Fukumura {\it et al.} \cite{TFEA} , even the out-of-plane anisotropy depends on Co content and carrier density. Thus, it is difficult to draw unified explanation of the magnetic anisotropy at present. 

In conclusion, we have studied  the high temperature ferromagnetism observed in rutile-type Ti$_{1-x}$Co$_x$O$_{2-\delta}$  films using x-ray magnetic circular dichroism at the Co $L_{2,3}$ edges (both in the TEY and TFY mode). These results represent that the high temperature ferromagnetism is originated from the Co$^{2+}$ atoms, most probably charge carriers induce the ferromagnetism. The magnetic moment of the Co ions as long as 0.82-2.25$\mu_B$/Co  was first observed by the bulk sensitive TFY method. The magnetic moment value deduced  with the TEY mode ( 0.15-0.24 $\mu_B$/Co) indicates the presence of a magnetically dead layer of $\sim$ 5 nm thickness on the sample surface. 

This work was supported by a Grant-in-Aid for Scientific Research in Priority Area ``Creation and Control of Spin Current'' (19048012) from MEXT, Japan and a Global COE Program ``the Physical Science Frontier'' from MEXT, Japan.


\begin{thebibliography}{references}
\bibitem{CMC} The calculated spectra are the weighted sum of 38 \%  of $D_{2h}$  low-spin, 38 \% of $O_{h}$  low spin and 24 \%  of $O_{h}$  high-spin crystal field. This result again suggests that the majority of the ferromagnetic Co ions in bulk Ti$_{1-x}$Co$_x$O$_2$ have Co$^{2+}$ with random crystal fields.
\end{thebibliography}
\end{document}